\documentclass[10pt,conference]{IEEEtran}
\usepackage{epsfig,setspace,amsmath,epsf,amssymb,bm,theorem,cite,graphicx, epstopdf,algorithm,algpseudocode,float,color,mathtools,authblk,tikz,physics}
\usepackage{optidef}
\usepackage{booktabs}
\usepackage{subfigure}
\usepackage{bbm}

\usetikzlibrary{positioning}
\usetikzlibrary{decorations.text}
\usetikzlibrary{decorations.pathmorphing}

\IEEEoverridecommandlockouts
\allowdisplaybreaks
\begin{document}

\title{Joint Age-State Belief is All You Need: \\ Minimizing AoII via Pull-Based Remote Estimation}
\author[1]{Ismail Cosandal}
\author[1]{Sennur Ulukus}
\author[2]{Nail Akar}

\affil[1]{\normalsize University of Maryland, College Park, MD, USA}
\affil[2]{\normalsize Bilkent University, Ankara, T\"{u}rkiye}

\maketitle
\begin{abstract}
Age of incorrect information (AoII) is a recently proposed freshness and mismatch metric that penalizes an incorrect estimation along with its duration. Therefore, keeping track of AoII requires the knowledge of both the source and estimation processes. In this paper, we consider a time-slotted pull-based remote estimation system under a sampling rate constraint where the information source is a general discrete-time Markov chain (DTMC) process. Moreover, packet transmission times from the source to the monitor are non-zero which disallows the monitor to have perfect information on the actual AoII process at any time. Hence, for this pull-based system, we propose the monitor to maintain a sufficient statistic called {\em belief} which stands for the joint distribution of the age and source processes to be obtained from the history of all observations. Using belief, we first propose a maximum a posteriori (MAP) estimator to be used at the monitor as opposed to existing martingale estimators in the literature. Second, we obtain the optimality equations from the belief-MDP (Markov decision process) formulation. Finally, we propose two belief-dependent policies one of which is based on deep reinforcement learning, and the other one is a threshold-based policy based on the instantaneous expected AoII. 
\end{abstract}

\section{Introduction}
Age of information (AoI) metric has recently been proposed to capture information freshness in remote estimation problems \cite{Yates__HowOftenShouldone}. The AoI metric quantifies information freshness by a monitor which keeps track of how long ago the latest received information packet in the system had been generated. However, it is argued in \cite{maatouk2020} that AoI may fall short of capturing freshness in certain estimation problems since it does not consider the dynamics of the sampled process since even though the latest received packet may have been generated a long time ago, it is possible that the source may not have changed since then, and therefore, the packet can still be fresh. Similarly, a recently received packet may contain stale information if the source has already changed its state after the packet was generated. 

Stemming from this drawback of AoI, \cite{maatouk2020} proposes an alternative freshness metric, namely age of incorrect information (AoII) that penalizes the mismatch between the source and its estimation over time, and regardless of when it is sampled, it defines the estimation as \emph{fresh} if it is the same as the source. Another interesting feature of AoII in contrast to AoI is that the monitor is not required to get a new sample to bring the age down to zero since the mismatch condition between the source and the monitor may as well be brought to end with a transition of the source to the estimated value at the monitor.

In this paper, we consider the following AoII minimization problem in which an information source observes a DTMC process, and a remote monitor estimates the process from the updates received by the monitor from the source. We consider a pull-based scheme such that the source transmits its current state whenever a pull request arrives at the source, and the transmission is completed in the next time slot. The monitor updates its estimation by considering the source dynamics with the MAP estimator. Notice that the monitor does not have full information on AoII including the very same time slot the most recent update is received. Additionally, we consider a sampling rate constraint on the monitor that limits the average number of pull requests it can send. Therefore, we aim to find an AoII-minimizing policy at the monitor with the timely generation of pull requests based on partial observations. 

Generally, AoII is investigated for symmetric Markov chains, and a single threshold policy is proposed to minimize the average AoII \cite{maatouk2020, chen2021minimizing, kriouile2022pull}. Additionally, in these works, the latest received information is used, termed as the \emph{martingale estimator} \cite{akar_ulukus_tcom24}, since it would be optimum only if the source were a martingale. On the other hand, in our previous works \cite{cosandal2024modelingC, cosandal2024aoiiC, cosandal2024multi}, we have shown that if the source process is a general asymmetric Markov chain, a simple threshold policy would not be guaranteed to perform optimally. More specifically, in \cite{cosandal2024aoiiC, cosandal2024multi} it is shown that the optimum transmission policy should take into account all the estimation, source, and age values. Similarly, it was shown in \cite{cosandal2024modelingC} that the monitor can reduce the average AoII value if the pull request rates are allowed to be dependent on the latest received information. However, in that work, we have considered a preemption mechanism at the transmitter which aborts the transmission of the current information packet if the source process changes before the packet is received. Similarly, the pull-based AoII minimization problem in \cite{kriouile2022pull} considers an immediate transmission that allows the monitor to keep track of AoII. On the other hand, in the system model here, the decision maker (the monitor) has no direct information on the source process and hence the current AoII. In this paper, a sufficient statistic corresponding to their joint probability distribution under the MAP estimation rule is derived. Therefore, this paper motivates to obtain policies based on this distribution.

The closest MDP formulation to our problem is the partially observable Markov decision process (POMDP) formulation that considers states which are not observable directly, but their probability distributions can be obtained with observations using partial information \cite{spaan2012partially}; this distribution is called \emph{belief}. The main assumption of this formulation is that an MDP can be defined from the unobserved states, and the expected reward of the system can be calculated from this distribution. Because of the curse of dimensionality of the underlying POMDP formulation, exact solutions \cite{smallwood1973optimal, kaelbling1998planning} for POMDPs are not tractable for problems with large state and action spaces. An approximate solution is the so-called \emph{myopic} policy which selects the action that minimizes the expected reward by ignoring its effects on future rewards \cite{krishnamurthy2016partially, zhao2008myopic}. In \cite{krishnamurthy2016partially}, it is shown that a myopic policy is optimum for POMDPs under certain conditions. Similarly, the Whittle index approach can be adapted for POMDPs to obtain an index policy \cite{meshram2021indexability, shao2021partially}.

Because of the dependency of the estimator on the belief, it is no longer straightforward to use the POMDP formulation for the problem of interest. However, we can formulate our problem as a \emph{belief-MDP} such that the belief in unobserved states is viewed as a fully-observable continuous-valued state of the belief-MDP, and its equivalence to a POMDP is shown in \cite{kaelbling1998planning}. In some cases \cite{hatami2023status, shao2021partially}, the belief-MDP can further be converted to an MDP with observable and finite states, and optimum policies can be obtained accordingly.

Deep reinforcement learning (DRL) has recently gained popularity for solving MDPs using the exploration-exploitation trade-off \cite{arulkumaran2017deep}. Indeed, the value function for a POMDP can be approximated from the belief using DRL \cite{egorov2015deep}, or from the observation \cite{jaakkola1994reinforcement}, and the action that minimizes the value function is applied as a sub-optimal policy.

POMDP formulation has been used to minimize AoI in several works \cite{shao2021partially, tahir2024collaborative, hatami2023status, stamatakis2021autonomous, liu2023optimizing, leng2019age, gong2020age}. In \cite{shao2021partially, tahir2024collaborative,gong2020age, liu2023optimizing}, system models involving multiple sensors and a single monitor have been studied for different scenarios. In these works, the monitor is only aware of the AoI of the sensor that successfully transmits at that time slot and estimates the AoI of other sensors based on its previous observations. In \cite{shao2021partially}, \cite{gong2020age} and \cite{liu2023optimizing}, sub-optimal policies are obtained via the index policy, the myopic policy, and the particle filter, respectively. On the other hand, the authors of \cite{liu2023optimizing} convert the POMDP into a fully-observable discrete space MDP problem, and obtain an optimum policy. A similar system model has been studied in \cite{chen2022uncertainty}. In that work, an entropy-based metric, namely \emph{uncertainty of information}, is minimized by employing an index policy. Additionally, the AoI minimization problems in \cite{stamatakis2021autonomous, leng2019age, hatami2023status} consider failure status, channel availability, and the battery level as unobservable states, respectively.
	
Under the assumption that each update includes a timestamp of the generation time, the monitor can access the correct AoI value for the time slot the update is received. On the other hand, when the delay on the channel is considered, the source process may change before the update arrives, thus no updates guarantee resetting AoII and the monitor never has the correct AoII value including the time slot an update arrives. Additionally, AoII metric depends on the dynamics of the source process, and it is upper bounded even if there is no sampling. These aspects make AoII problems different and more challenging from AoI-based formulations.  

The contribution of this paper can be summarized as follows: i) We propose a MAP estimator to be used at the monitor in place of the simple martingale estimator, for which the monitor updates its estimation with the MAP rule. ii) We derive a sufficient statistic, namely belief, that corresponds to the joint distribution of AoII and source state, for general asymmetric Markov chains. iii) We propose two belief-dependent policies, one of which is based on DRL, and we compare these two policies against two baseline belief-agnostic policies.

\section{System Model}
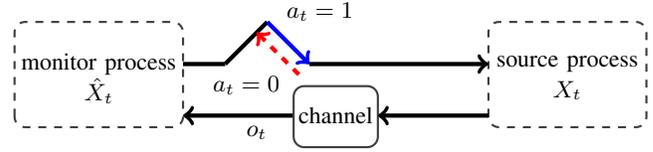
\begin{figure}[t]
    \centering
    \begin{tikzpicture}[scale=0.28]
    \draw[rounded corners,thick,darkgray,dashed] (-1,-1) rectangle (7,4) {};
    \filldraw (3,2) node[anchor=center] {\small{monitor process}};
    \filldraw (3,0.75) node[anchor=center] {\small{$\hat X_t$}};
    \draw[rounded corners,thick,darkgray,dashed] (21.5,-1) rectangle (29,4) {};
    \filldraw (25.25,2) node[anchor=center] {\small{source process}};
    \filldraw (25.25,0.75) node[anchor=center] {\small{${X}_t$}};
    \filldraw (10.5,-0.5) node[anchor=north] {\small{$o_t$}};
    \draw[ultra thick] (7,2) -- (9,2);
    \draw[ultra thick] (9,2) -- (11,4);
    \draw[->,blue,ultra thick] (11,4) -- (13,2);
    \node at (11,2) (nodeA) {};
    \node at (16,7) (nodeB) {};
    \draw[ultra thick,->] (13,2) -- (21.5,2);
    \filldraw (13.5,4.5) node[anchor=center] {\small{$a_t=1$}};
    \draw[->,red,ultra thick,dashed] (12.5,1.5) -- (10.5,3.5);
    \node at (8,-1) (nodeA) {};
    \node at (12,3) (nodeB) {};
    \filldraw (10,1) node[anchor=center] {\small{$a_t=0$}};
    \draw[ultra thick,<-] (7,-0.45) -- (12.25,-0.45);
    \draw[rounded corners,thick,darkgray] (12.25,-1.95) rectangle (16.25,.95) {};
    \filldraw (14.25,-0.45) node[anchor=center] {\small{channel}};
    \draw[ultra thick,<-] (16.25,-0.45) -- (21.5,-0.45);
    \end{tikzpicture}
	\caption{Pull-based transmission model between a source and a monitor.}
    \label{fig:sys}
\end{figure}

Consider a time-slotted transmission model between a source and a monitor; see Fig.~\ref{fig:sys}. The source observes a finite discrete-time Markov chain process with $N$ states, denoted by $X_t$, that stays at the state during a time slot, and then a state transition occurs according to a state transition matrix $\bm{P}$. 

We consider a pull-based transmission scheme for which the monitor takes an action $a_t$ at the start of the time slot $t$ to send a pull request ($a_t=1$) or not ($a_t=0$).
Pull requests are assumed to arrive instantaneously at the source. Upon receiving the pull request, the source samples the current process and transmits the information packet to the monitor. We consider a one-way delay channel between them; pull requests reach the sources immediately, but transmission from the source to the monitor is completed after one slot.

The monitor obtains an observation $o_t \in \mathcal{O}(a_{t-1})$ from the observation spaces $\mathcal{O}(0)=\emptyset$ and $\mathcal{O}(1)=\{1,2,\dots,N\}$ as,
\begin{align}
    o_t=\begin{cases}
        X_{t-1}, & a_{t-1}=1, \\
        \emptyset, & a_{t-1}=0.
    \end{cases} \label{eq:obs}
\end{align}

By using this observation, the monitor estimates the current state of the source process as a vector $\bm{\pi}_t=[\pi_t(1),\dots,\pi_t(N)]$ where $\pi_t(k)=\mathbb{P}(X_t=k|o_{u(t)})$ where $u(t)$ is the generation time of the latest received message, equivalently, time of the latest pull request, i.e.,
\begin{align}
    u(t)=\max(\eta | a_{\eta }=1,\eta \leq t).
\end{align}
Thus, the evolution of $\pi$ can be expressed as,
\begin{align}
    \bm{\pi}_t=\bm{e}_{o_{u(t)}}\bm{P}^{t-u(t)},
\end{align}
where $\bm{e}_j$ is a row vector of zeros except for a one in the $j$th position. Additionally, the monitor employs the MAP estimator $\hat{X}_t$, which can be expressed in terms of $\pi_t$ as,
\begin{align}
 \hat{X}_t=\arg\max \bm\pi_t.  
\end{align}

\begin{figure}[t]
    \centering
    \includegraphics[width=0.98\linewidth]{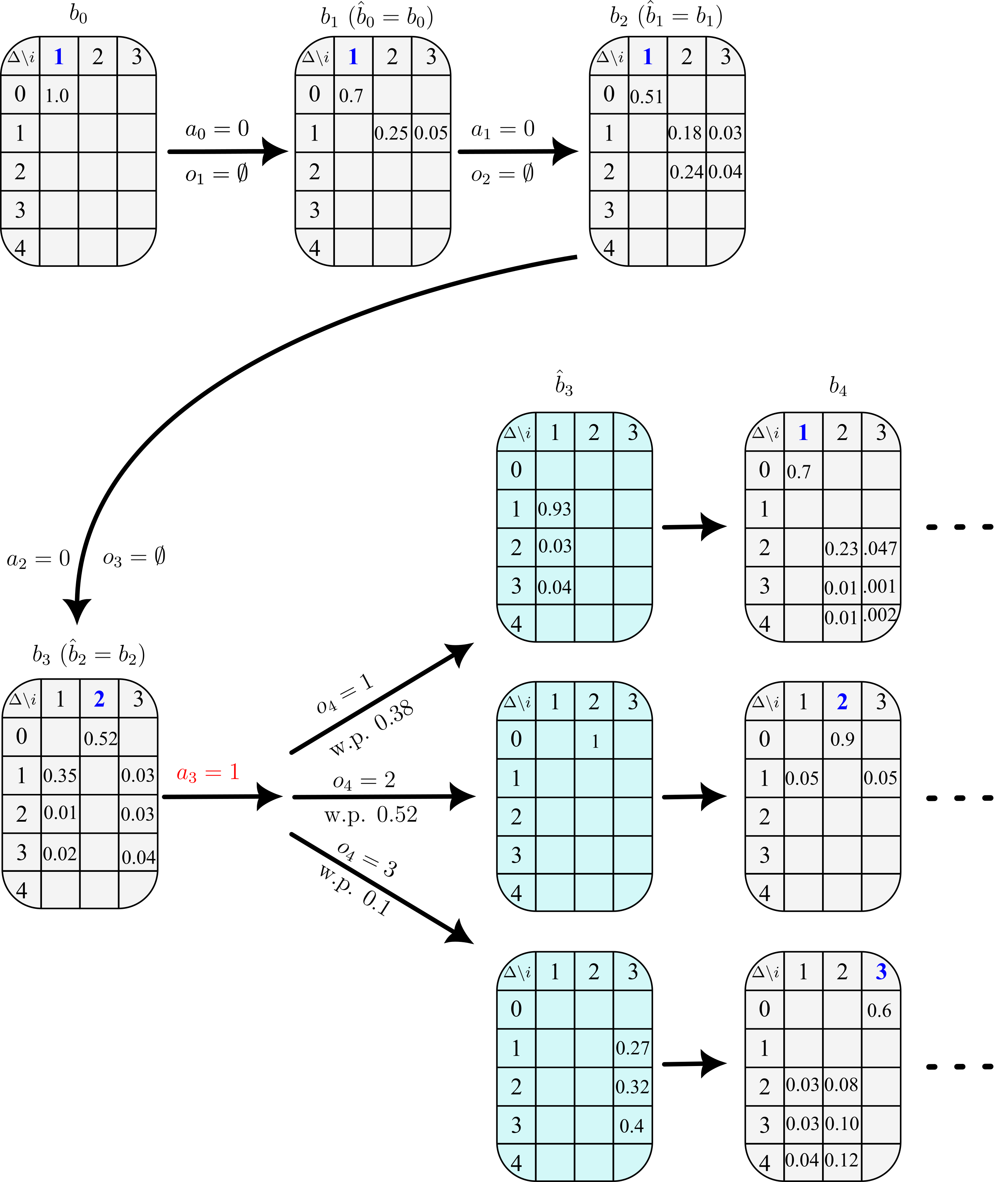}
    \caption{Evolution of the belief for the ternary source whose transition probability matrix $\bm{P}_2$ is given in \eqref{sim-result-matrices} in the simulations result section. The process starts from state 1. At each time slot, the monitor estimates the state with the highest likelihood, which is highlighted in blue. For the first two time slots, the monitor stays idle, and at the third time slot, the monitor sends a pull request. After receiving an observation at time slot 4, the monitor first updates the belief at time slot 3 as $\hat{b}_3$ accordingly, and subsequently obtains the belief matrix at time slot 4 by using the update. All possible observations are illustrated with their relevant probabilities.}
    \label{fig2}
\end{figure}

The mismatch between $X_t$ and $\hat{X}_t$ is measured by using the AoII metric. The AoII metric penalizes the estimation error linearly while the error stays, and it is reset to zero when $X_t$ and $\hat{X}_t$ are synchronized. Defining $\text{AoII}_{t}$ to be the value of AoII at time $t$, its evolution can be expressed as,
\begin{align}
\text{AoII}_{t+1}=\begin{cases}
    \text{AoII}_{t}+1, & X_t\neq \hat{X}_t, \\
    0, & X_t=\hat{X}_t.
\end{cases} \label{eq:aoii}
\end{align} 

Since the monitor will never have the actual state of the source at any time due to the communication delay, it also would not know the instantaneous value of AoII$_t$. Instead, the monitor estimates its distribution by using all observations. In fact, the distribution of AoII is not independent of the source state. Thus, we define the joint belief as,
\begin{align}
    b_t(i,\Delta)=\mathbb{P}(X_t=i, \text{AoII}_t=\Delta | H_t), \label{eq:b}
\end{align}
where $H_t=\{o_1,o_2,\dots,o_t\}$ is defined as the history of all observations until time $t$. The monitor calculates this belief in two steps as described below and exemplified in Fig.~\ref{fig2}. 

First, assume that the monitor has the belief $b_{t-1}$ at the beginning of time slot $t$, and a new observation $o_t$ arrives. Notice from \eqref{eq:obs} that if $a_{t-1}=1$, the new observation includes information about $X(t-1)$, thus the belief should be updated accordingly. We define the updated belief as $\hat{b}_t(i,\Delta)=\mathbb{P}(X_t=i, \text{AoII}_t=\Delta|H_{t+1})$ and the update can be performed as,
\begin{align}
\hat{b}_{t-1}(i,\Delta)=\begin{cases}
    b_{t-1}(i,\Delta), &  o_t=\emptyset, \\
    \dfrac{b_{t-1}(i,\Delta)}{\sum_{j=1}^N b_{t-1}(j,\Delta)}, & o_t=i, \\
    0, & o_t= j, \ j\neq i.  \label{eq:hat_b_t}
\end{cases}    
\end{align}
Then, the belief for time slot $t$ can be calculated using the updated belief, and the source dynamics as follows,
\begin{align}
b_t(i,\Delta)=\begin{cases}
        \max  \bm{\pi}_t, &  i=\hat{X}_t, \ \Delta=0, \\
        \sum_{m=1}^{N}\hat{b}_{t-1}(m,\Delta-1)\bm{P}_{mi}, &  i\neq\hat{X}_t, \ \Delta>0,\\
        0, & \text{otherwise}.
    \end{cases} \label{eq:b_t}
\end{align}

We remind that $\bm{\pi}_t$ is a function of $b_t$ since $\pi_t(k)=\sum_{\Delta}b_t(k,\Delta)$.
It is a well-known result that belief $b_t$ includes all necessary information about the history $H_t$ regarding decision making, and therefore, it is a sufficient statistic \cite{smallwood1973optimal}. We can justify this from the following identity that results from \eqref{eq:hat_b_t} and \eqref{eq:b_t},
\begin{align}
    \mathbb{P}(b_{t+1}|H_{t+1})=\mathbb{P}(b_{t+1}|o_{t+1},{b}_{t})=\mathbb{P}(b_{t+1}|\hat{b}_{t}).
\end{align}

Finally, we define transition probabilities for joint state-AoII beliefs for action $a_t$ as $T(b_{t+1},a_t,b_t)=\mathbb{P}(b_{t+1}|a_t,b_t)$. This probability can be expressed as,  
\begin{align}
 T(b_{t+1},a_t,b_t)&=\mathbb{P}(b_{t+1}|a_t,b_t) \\
 &= \mathbb{P}(b_{t+1}|a_t,b_t,o_{t+1})\mathbb{P}(o_{t+1}|a_t,b_t) \label{eq:Tb}
\end{align}
and it can be obtained using \eqref{eq:hat_b_t} and \eqref{eq:b_t}, and conditional probability distribution of observation, which is,
\begin{align}
    \mathbb P(o_t=k|a_{t-1}=1,b_{t-1})=&\pi_{t-1}(k), \\
    \mathbb P(o_t=\emptyset|a_{t-1}=0,b_{t-1})=&1.
\end{align}

\section{Problem Formulation}
We consider a sampling constraint on the monitor such that it cannot send pull requests every time slot. With a budget $\alpha$ for the average sampling rate, the constrained optimization problem is stated as,
\begin{mini}
	{\phi}{\text{MAoII}^{\phi}} 
	{\label{Opt1}}
    {}
	\addConstraint{ R^{\phi} }{\leq \alpha,} 
\end{mini}
where $\text{MAoII}$ denotes the time average of AoII, i.e., $\text{MAoII}= \lim_{T\to\infty} \frac{1}{T} \sum_{t=0}^T \text{AoII}(t)$, $R$ denotes the average sampling rate as $R=\lim_{T\to\infty} \frac{1}{T} \sum_{t=0}^T a_t$, and $\text{MAoII}^{\phi}$ (resp. $R^{\phi}$) denotes the average AoII (resp. sampling rate) obtained by imposing policy $\phi$. This problem can be converted to an unconstrained problem with a Lagrangian coefficient as,
\begin{mini}
	{\phi}{\text{MAoII}^{\phi}+\lambda R^{\phi}.} 
	{\label{OptUC}}
    {}
\end{mini}

It is known that \cite{makowski1986} if there exists a Lagrangian coefficient $\lambda^*$ such that the optimum policy $\phi^*$ obtained from the unconstrained problem is also optimum for the constrained problem either when (i) the constrained problem attains $R^{\phi^*}=\alpha$, or (ii) $\lambda=0$ and $R^{\phi^*}\leq \alpha$. However, from the nature of the discrete-time system, a deterministic policy on the boundary of the constraint set may not exist. For such cases, a mixture of multiple deterministic policies can be used to obtain an optimal policy for the unconstrained problem. Among many methods, we adopt a non-randomized past-dependent policy,, namely, \emph{steering algorithm} in \cite{ross1989randomized}. In the steering algorithm, we consider two policies $\phi_-$ and $\phi_+$ such that $R^{\phi_-}\leq\alpha\leq R^{\phi_+}$, and the algorithm switches between the two policies based on the current sampling rate. The general procedure of the algorithm is summarized in Algortihm~\ref{alg:steer}.

\begin{algorithm}[h]
    \caption{Steering algorithm}\label{alg:steer}
    \begin{algorithmic}
    \State \textbf{Input:} $\phi_-$ and $\phi_+$ such that $R^{\phi_-}\leq\alpha\leq R^{\phi_+}$
    \State \textbf{Initialize:} $R_0=0$, $N=0$
        \For{$t \gets 1$ to $T$}                    
    \If{$R_{t-1}< b$}
    \State Obtain $a_t$ by applying the policy $\phi_+$ 
    \Else
    \State Obtain $a_t$ by applying the policy $\phi_-$
    \EndIf
    \State{$N \gets N + a_t$}
    \State{$R_t \gets \frac{N}{t}$}
    \EndFor     
    \end{algorithmic}
\end{algorithm}

Let us define the unobserved states of the problem as  $s_t=(i,\Delta)\in \mathcal{S}$, where $\mathcal{S}$ is defined $ \{\{1,\dots,N\} \times \{0,\dots,\Delta_{\max}\} \}$. Note that for practical reasons, we have truncated age values to $\Delta_{\max}$. Additionally, because of MAP estimation, the unobserved state AoII depends on the estimation, and the estimation depends on the belief. Therefore, it is not possible to obtain transition probabilities between these states from POMDP formulation.  On the other hand, we can obtain Bellman's equations for the optimization problem in \eqref{OptUC} as an equivalent belief-MDP($\mathcal{B},\mathcal{A},T,r^{\lambda},\gamma$) in \cite{kaelbling1998planning}:
\begin{itemize}
    \item As defined in \eqref{eq:b}, $b_t$ includes $N(\Delta_{max}+1)$ elements, each representing a probability. Thus, the belief space can be defined as $\mathcal{B}=[0,1]^{ N(\Delta_{max}+1)}$. 
    \item Action space is defined $\mathcal{A}=\{0,1\}$, and each action $a_t \in \mathcal{A}$ is a feasible action for any state.
    \item $T(b_{t+1},a_t,b_t)$ is the transition probability between belief states, and it is defined in \eqref{eq:Tb}.
    \item $r^{\lambda}(b_t,a_t)$ is the reward of the problem in \eqref{OptUC}. For the unobserved state, it is equal to $r^{\lambda}(s_t=\{i,\Delta\},a_t)=\Delta+\lambda a_t$. Thus, for a given $b_t$ and $a_t$ pair, it can be obtained as,
    \begin{align}
     r^{\lambda}(b_t,a_t)=&\sum_{i=1}^N \sum_{\Delta=0}^{\Delta_{\max}}  r^{\lambda}(s_t=\{i,\Delta\},a_t)b_t(i,\Delta) \\
    =&\sum_{\Delta=0}^{\Delta_{max}}\Delta\sum_{i=1}^N b_t(i,\Delta)+a_t\lambda.
    \end{align}
    Notice from \eqref{eq:b} that $r^\lambda(b_t,a_t)$ is equivalent to,
    \begin{align}
     r^\lambda(b_t,a_t)=&\sum_{i=1}^N\sum_{\Delta=0}^{\Delta_{max}}(\Delta+a_t\lambda)\mathbb{P}(X_t=i,\text{AoII}_t=\Delta|H_t)\\=&\mathbb{E}[\text{AoII}_t|H_t]+a_t\lambda.
    \end{align}
    Additionally, we note that regardless of $\lambda$, the instantaneous reward when $a_t=0$ is equal to expected AoII, and we denote this as,
    \begin{align}
     r^\lambda(b_t,0)=r(b_t,0)=\mathbb{E}[\text{AoII}_t|H_t].
    \end{align}
    \item $\gamma$ is the discount factor.
\end{itemize}

Finally, Bellman's optimality equation \cite{spaan2012partially} for the unconstrained problem in \eqref{OptUC} for any Lagrangian coefficient $\lambda$ can be expressed as,
\begin{align}
    &V^\lambda(b_t,a_t) \nonumber\\
    &=r^\lambda(b_t,a_t) + \gamma\sum_{o \in \mathcal{O}(a)} T(b_t^o,a_t,b_t)  \min_{a'\in\{0,1\}}V(b_t^o,a'), \label{eq:bell}
\end{align}
where $V^\lambda(b_t,a_t)$ is the average cost attained when the action $a_t$ is applied at initial belief $b_t$ first, but then the optimum policy is applied, and 
\begin{align}
b^o=\{b_{t+1}|b_t=b,o_{t+1}=o\}    \label{eq:b_o}
\end{align}
corresponds to the new belief evaluated from $b$ under the observation $o$.

\section{Proposed Solutions}
Since belief space includes uncountably many elements, we cannot obtain the value function for each belief with dynamic programming from the belief-MDP representation. Therefore, next, we will discuss two suboptimal policies: i) a policy obtained with non-linear approximation by deep reinforcement learning, and ii) a greedy policy that a pull request is sent if the expected immediate reward exceeds a threshold.

\subsection{Non-Linear Value Function Approximation With DRL}
\begin{figure}[!t]
    \centering
    \includegraphics[width=0.85\linewidth]{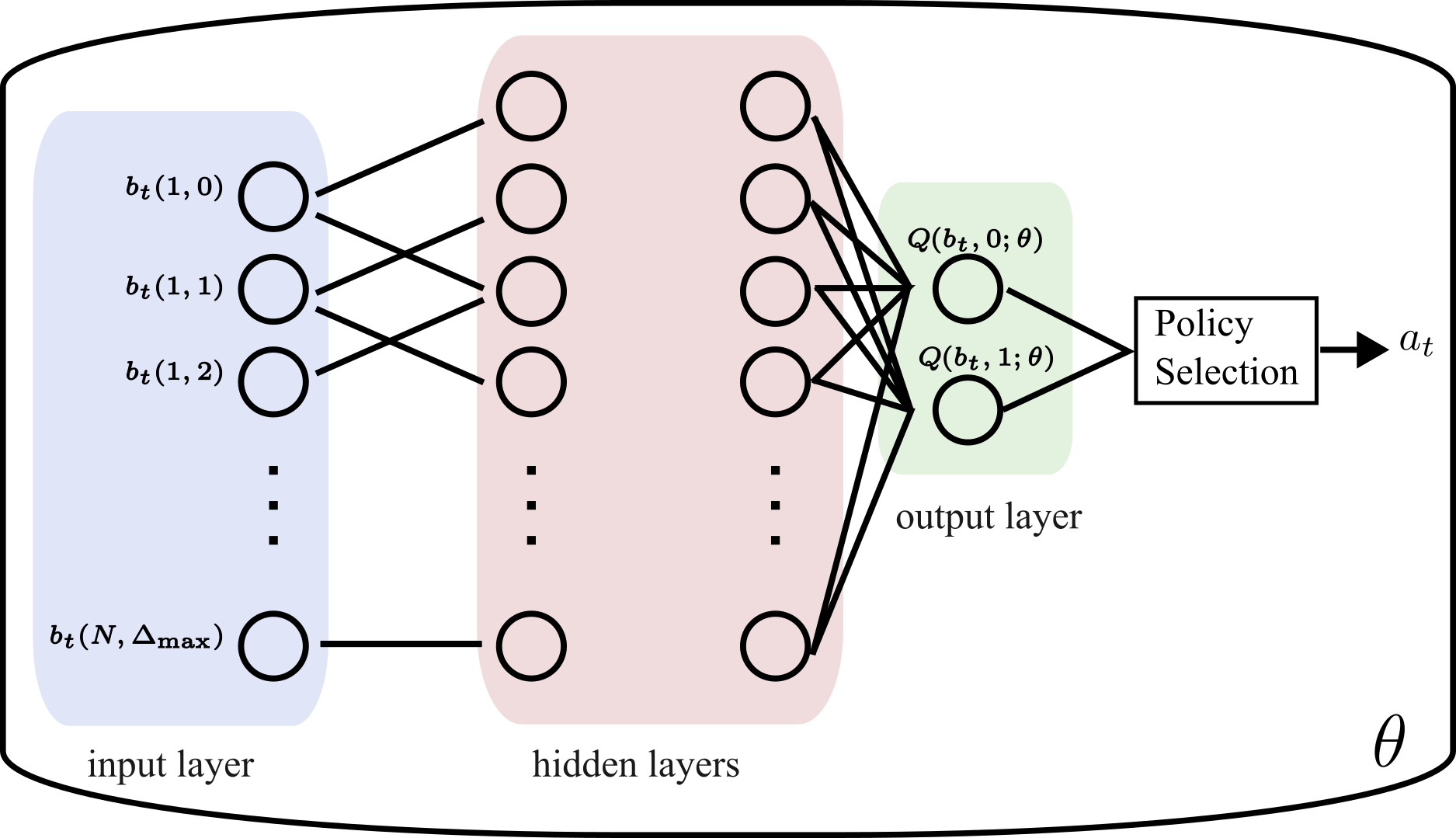}
    \caption{Network architecture used for DRL with parameter $\theta$.}
    \label{fig:RL}
\end{figure}

In order to find a sub-optimal policy, first we use a non-linear approximation for the value function in \eqref{eq:bell}. The network architecture includes fully connected layers and a policy selection step as summarized in Fig.~\ref{fig:RL}. To improve the convergence of learning \cite{egorov2015deep}, we construct two networks, namely, a main network, and a target network with parameters $\theta$, and $\theta'$, respectively. Target network parameters are frozen for $L-1$ steps, and then equalized as $\theta'\leftarrow \theta$ periodically on each $L$th step. Main network parameters $\theta$, on the other hand, are obtained to minimize the cost function,
\begin{align}
        J(\theta)=&Q^\lambda(b_t,a_t;\theta)-\Bigg(r^\lambda(b_t,a_t) \nonumber
        \\&+ \gamma\sum_{o \in \mathcal{O}(a)} T(b^o,a_t,b_t) \min_{a'\in\{0,1\}}Q^\lambda(b^o,a';\theta')\Bigg), \label{eq:costdqn}
\end{align}
where Q-values $Q^\lambda(b_t,a_t;\theta)$ give the estimated cost for the belief $b_t$, and an action $a_t$ for the policy obtained from the network is applied.

Throughout the learning phase, the policy is selected via exploration-exploitation with an exploration coefficient $0<\delta<1$, and a decaying coefficient $\nu\approx 1$ for $e$ is the epoch number,
\begin{align}
    a_t=\begin{cases}
        \arg\min Q^\lambda(b_t,a_t;\theta), & \text{w.p.} \quad 1-\delta\cdot \nu^{e-1}, \\
        $0$, & \text{w.p.} \quad \dfrac{\delta\cdot \nu^{e-1}}{2}, \\
        $1$, & \text{w.p.} \quad \dfrac{\delta\cdot \nu^{e-1}}{2}. 
    \end{cases}
\end{align}
After the network parameters converge, or $e$ reaches the maximum epoch number $e_{\max}$, the learning phase is finalized, and the policy $\phi_{\lambda}$ is obtained as,
\begin{align}
    a_t= \arg\min Q^\lambda(b_t,a_t;\theta).
\end{align}

It is possible for the algorithm to get stuck at a local minima, thus, the learning step is repeated multiple times for the same $\lambda$ value, and the policy that gives the minimum cost value is selected. Then, the mixture policy for Lagrangian values $\lambda^+=\text{inf}\{\lambda| \alpha\geq R^{\phi_\lambda}\}$ and $\lambda^-=\text{sup}\{\lambda| \alpha\leq R^{\phi_\lambda}\}$ is obtained via the steering algorithm in Algorithm~\ref{alg:steer}. 

\subsection{Expected AoII Threshold Policy}
From \eqref{eq:b_t} one can observe that for any ergodic process $P$, the belief process under no observations is also ergodic, and it has a steady state $b^{st}$. Thus, the expected AoII, which is also equivalent to immediate reward under $a_t=0$, reaches a steady-state value $r(b^{st},0)$ if no pull request is sent. From this observation, the second belief-dependent policy we propose is a threshold policy in which the monitor sends the pull request only if the expected AoII exceeds a certain value. For any threshold value $\tau$, the policy $\phi_\tau$ is given as follows,
\begin{align}
    a_t=\begin{cases}
        1, &  r(b_t,0)\geq \tau, \\
        0, &  r(b_t,0) < \tau. 
    \end{cases} \label{eq:ind_pol}
\end{align}

Unlike DRL, this policy directly attempts to solve the constrained problem in \eqref{Opt1}. MAoII and $R$ values for each candidate threshold $\tau$ are obtained via a simulation. Candidate threshold values $\tau^-$ and $\tau^+$ that satisfy condition $R^{\phi_{\tau^-}}\leq\alpha\leq R^{\phi_{\tau^+}}$ can be found with a bisection search, and a mixture of these two policies is obtained via the steering algorithm in Algorithm~\ref{alg:steer}. 

\section{Simulation Results}
To compare our proposed belief-dependent policies, we use two belief-agnostic policies as benchmarks, namely, \emph{random policy} and \emph{uniform policy}. In the random policy, in each time slot, the monitor sends a pull request with $\alpha$ probability. In the uniform policy, on the other hand, pull requests are generated almost uniformly with a period ${1}/{\alpha}$. In other words, the $m$th pull request is generated at $\text{round}({m}/{\alpha})$th time slot, where $\text{round}(\cdot)$ operation rounds the inputs to the nearest integer. In \cite{Yates__HowOftenShouldone}, it is proven that if the service time is fixed, selecting the inter-generation times uniform minimizes the AoI, thus the uniform sampling is the optimum policy for AoI minimization. Notice that both policies satisfy the sampling constraint at the boundary, i.e., with equality.

\begin{table}
    \centering
    \caption{Parameters used in the simulation results.}
    \begin{tabular}[t]{|c|c|c|}
    \hline
    $\Delta_{\max}$& 15\\ \hline
    $T$ & $10^5$ \\ \hline
    learning rate& $10^{-3}$\\ \hline
    numbers of hidden layers& 2\\ \hline
    nodes at hidden layers& $60$\\ \hline
    \end{tabular}
    \begin{tabular}[t]{|c|c|c|}
    \hline
    $L$& $50$\\ \hline
    $e_{\max}$& 50\\ \hline
    $\nu$& 0.9\\ \hline
    $\gamma$& $0.95$\\ \hline
    $\delta$  & $.25$ or $.05$\\  \hline
    \end{tabular}
    \label{tab:sim_tab}
\end{table}

In the simulations, we consider a binary and a ternary source with transmission matrices,
\begin{align}
    \bm{P}_1=\begin{bmatrix}
    0.85 & 0.15 \\ 
    0.25 & 0.75
    \end{bmatrix}, \quad 
    \bm{P}_2=\begin{bmatrix} 
    0.70 & 0.25 & 0.05 \\
    0.05 & 0.90 & 0.05 \\
    0.10 & 0.30 & 0.60
    \end{bmatrix},
    \label{sim-result-matrices}
\end{align}
and the remaining parameters are summarized in Table~\ref{tab:sim_tab}.

\begin{figure}
    \centering
    \includegraphics[width=0.95\linewidth]{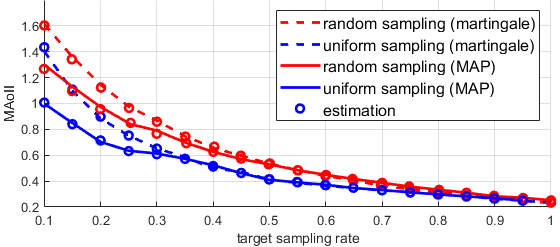}
    \caption{Comparing the martingale and MAP estimators with belief-agnostic policies for the binary process $\bm{P}_1$.}
    \label{fig:mart_map}
\end{figure}

\begin{figure}[t]
    \begin{center}
    \subfigure[]{\includegraphics[width=0.48\linewidth]{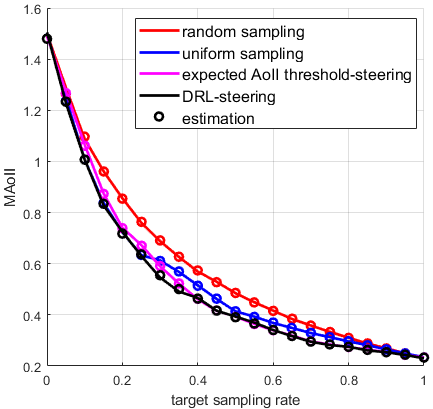}}
    \subfigure[]{\includegraphics[width=0.48\linewidth]{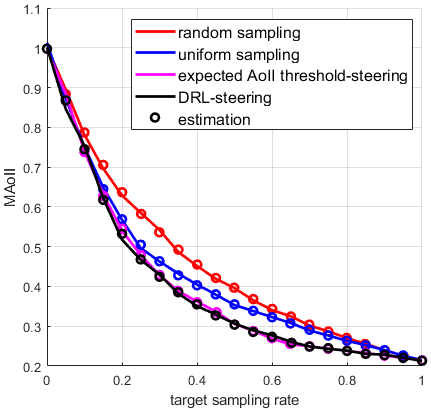}} 
    \end{center}  
    \centering    
    \caption{Comparing the proposed algorithm with age-and-estimation agnostic transmission policy for (a) binary source $\bm{P}_1$, and (b) ternary source $\bm{P}_2$.}
    \label{fig:vs}
\end{figure}

In all simulation results, the lines correspond to the real MAoII values, and the circles correspond to the estimates of MAoII obtained from the belief as $\hat{\Delta}=\frac{1}{T}\sum_{t=0}^T r(b_t,0)$. First, from the consistency between the estimates and the real values of MAoII in all simulation results in Figs.~\ref{fig:mart_map} and \ref{fig:vs}, we thus have verified our calculations for the belief.

In Fig.~\ref{fig:mart_map}, we illustrate the effect of the estimator on benchmark policies for the process $\bm{P}_1$. In general, we can conclude that the MAP estimator is superior to the martingale estimator. Notice that the difference between the MAP and martingale estimators disappears for larger sampling rates, because when there is frequent sampling, there would be no need to update the estimator. Notice that unlike the MAP estimator, the martingale estimator is not defined for the target sampling rate $0$. 

Finally, we compare all policies with Fig.~\ref{fig:vs}(a) and Fig.~\ref{fig:vs}(b) for processes $\bm{P}_1$ and $\bm{P}_2$, respectively. These figures indicate that belief-dependent policies, specifically DRL policy, outperform belief-agnostic policies in all cases. Additionally, despite its low complexity, the expected AoII policy performs similarly to the DRL in most cases. Notice that in low sampling rates, uniform sampling has similar performance to the belief-dependent policies. As discussed earlier, if no actions are taken for a long period of time, the belief reaches steady-state $b^{st}$, and the estimation becomes $\hat{X}=\arg\min b^{st}$, from the MAP estimator. In other words, uniform sampling sends pull requests when the belief belongs to the same sampling space that is similar to what belief-dependent policies do.

\section{Conclusion}
We considered a system model in which the monitor has imperfect information on the source's state and therefore on the AoII process at all times, however, a sufficient statistic \emph{belief} can be obtained from the observations. We proposed a MAP estimation rule as a function of the belief, and two control policies based on this estimation rule and the belief so as to minimize the average AoII. We showed that the MAP estimator improves the average AoII in comparison to the martingale estimator used in most existing studies. Moreover, sending belief-dependent pull requests further improves the average AoII. The proposed scheme outperforms a pair of belief-agnostic policies used as benchmark for comparison in this paper. The proposed belief-dependent control approach has the potential to be used in other related information freshness problems.

\bibliographystyle{unsrt}
\bibliography{bibl}

\begin{thebibliography}{10}

\bibitem{Yates__HowOftenShouldone}
S.~Kaul, R.~Yates, and M.~Gruteser.
\newblock Real-time status: How often should one update?
\newblock In {\em IEEE Infocom}, March 2012.

\bibitem{maatouk2020}
A.~Maatouk, S.~Kriouile, M.~Assaad, and A.~Ephremides.
\newblock The age of incorrect information: A new performance metric for status updates.
\newblock {\em IEEE/ACM Trans. Netw.}, 28(5):2215--2228, October 2020.

\bibitem{chen2021minimizing}
Y.~Chen and A.~Ephremides.
\newblock Minimizing age of incorrect information for unreliable channel with power constraint.
\newblock In {\em IEEE Globecom}, December 2021.

\bibitem{kriouile2022pull}
S.~Kriouile and M.~Assaad.
\newblock When to pull data from sensors for minimum distance-based age of incorrect information metric.
\newblock In {\em IEEE WiOpt}, February 2022.

\bibitem{akar_ulukus_tcom24}
N.~Akar and S.~Ulukus.
\newblock Query-based sampling of heterogeneous {CTMC}s: Modeling and optimization with binary freshness.
\newblock {\em IEEE Trans. Comm.}, 2024.
\newblock To appear.

\bibitem{cosandal2024modelingC}
I.~Cosandal, N.~Akar, and S.~Ulukus.
\newblock Modeling {AoII} in push- and pull-based sampling of continuous time {M}arkov chains.
\newblock In {\em IEEE Infocom}, May 2024.

\bibitem{cosandal2024aoiiC}
I.~Cosandal, N.~Akar, and S.~Ulukus.
\newblock {AoII}-optimum sampling of {CTMC} information sources under sampling rate constraints.
\newblock In {\em IEEE ISIT}, July 2024.

\bibitem{cosandal2024multi}
I.~Cosandal, N.~Akar, and S.~Ulukus.
\newblock Multi-threshold {AoII}-optimum sampling policies for {CTMC} information sources.
\newblock 2024.
\newblock Available online at arXiv:2407.08592.

\bibitem{spaan2012partially}
M.~T.~J. Spaan.
\newblock Partially observable {M}arkov decision processes.
\newblock In {\em Reinforcement Learning: State-of-the-Art}, pages 387--414. Springer, 2012.

\bibitem{smallwood1973optimal}
R.~D. Smallwood and E.~J. Sondik.
\newblock The optimal control of partially observable {Markov} processes over a finite horizon.
\newblock {\em Operations Research}, 21(5):1071--1088, September 1973.

\bibitem{kaelbling1998planning}
L.~P. Kaelbling, M.~L. Littman, and A.~R. Cassandra.
\newblock Planning and acting in partially observable stochastic domains.
\newblock {\em Artificial intelligence}, 101(1-2):99--134, May 1998.

\bibitem{krishnamurthy2016partially}
V.~Krishnamurthy.
\newblock {\em Partially Observed Markov Decision Processes}.
\newblock Cambridge University Press, 2016.

\bibitem{zhao2008myopic}
Q.~Zhao, B.~Krishnamachari, and K.~Liu.
\newblock On myopic sensing for multi-channel opportunistic access: {S}tructure, optimality, and performance.
\newblock {\em IEEE Trans. Wirel. Commun.}, 7(12):5431--5440, December 2008.

\bibitem{meshram2021indexability}
R.~Meshram and K.~Kaza.
\newblock Indexability and rollout policy for multi-state partially observable restless bandits.
\newblock In {\em IEEE CDC}, December 2021.

\bibitem{shao2021partially}
Y.~Shao, Q.~Cao, S.~C. Liew, and H.~Chen.
\newblock Partially observable minimum-age scheduling: The greedy policy.
\newblock {\em IEEE Trans. Comm.}, 70(1):404--418, October 2021.

\bibitem{hatami2023status}
M.~Hatami, M.~Leinonen, and M.~Codreanu.
\newblock Status updating under partial battery knowledge in energy harvesting {IoT} networks.
\newblock 2023.
\newblock Available online at arXiv:2303.18104.

\bibitem{arulkumaran2017deep}
K.~Arulkumaran, M.~P. Deisenroth, M.~Brundage, and A.~A. Bharath.
\newblock Deep reinforcement learning: A brief survey.
\newblock {\em IEEE Signal Process. Mag.}, 34(6):26--38, November 2017.

\bibitem{egorov2015deep}
M.~Egorov.
\newblock Deep reinforcement learning with {POMDP}s.
\newblock {\em Tech. Rep.}, December 2015.

\bibitem{jaakkola1994reinforcement}
T.~Jaakkola, S.~Singh, and M.~Jordan.
\newblock Reinforcement learning algorithm for partially observable {Markov} decision problems.
\newblock {\em Adv. Neural. Inf. Process. Syst.}, 7, 1994.

\bibitem{tahir2024collaborative}
A.~Tahir, K.~Cui, B.~Alt, A.~Rizk, and H.~Koeppl.
\newblock Collaborative optimization of the age of information under partial observability.
\newblock In {\em IFIP Networking}, August 2024.

\bibitem{stamatakis2021autonomous}
G.~Stamatakis, N.~Pappas, A.~Fragkiadakis, and A.~Traganitis.
\newblock Autonomous maintenance in {IoT} networks via {AoI}-driven deep reinforcement learning.
\newblock In {\em IEEE Infocom}, May 2021.

\bibitem{liu2023optimizing}
J.~Liu, Q.~Wang, and H.~H. Chen.
\newblock Optimizing age of information in uplink multiuser {MIMO} networks with partial observations.
\newblock In {\em IEEE WiOpt}, August 2023.

\bibitem{leng2019age}
S.~Leng and A.~Yener.
\newblock Age of information minimization for an energy harvesting cognitive radio.
\newblock {\em IEEE Trans. Cogn. Commun. Netw.}, 5(2):427--439, May 2019.

\bibitem{gong2020age}
A.~Gong, T.~Zhang, H.~Chen, and Y.~Zhang.
\newblock Age-of-information-based scheduling in multiuser uplinks with stochastic arrivals: A {POMDP} approach.
\newblock In {\em IEEE Globecom}, December 2020.

\bibitem{chen2022uncertainty}
G.~Chen, S.~C. Liew, and Y.~Shao.
\newblock Uncertainty-of-information scheduling: A restless multiarmed bandit framework.
\newblock {\em IEEE Trans. Inf. Theory}, 68(9):6151--6173, August 2022.

\bibitem{makowski1986}
D.~J. Ma, A.~M. Makowski, and A.~Shwartz.
\newblock Estimation and optimal control for constrained {Markov} chains.
\newblock In {\em IEEE CDC}, December 1986.

\bibitem{ross1989randomized}
K.~W. Ross.
\newblock Randomized and past-dependent policies for {Markov} decision processes with multiple constraints.
\newblock {\em Operations Research}, 37(3):474--477, May 1989.

\end{thebibliography}

\end{document}